\documentstyle[amssymb,floats,aps,prb,epsf]{revtex}
\epsfclipon
\begin{document}
\twocolumn[\hsize\textwidth\columnwidth\hsize\csname
@twocolumnfalse\endcsname

\draft
\title{Interplay between single particle coherence and
kinetic energy driven superconductivity in doped cuprates}
\author{Tianxing Ma, Huaiming Guo, and Shiping Feng$^{*}$}
\address{Department of Physics, Beijing Normal University, Beijing
100875, China}
\maketitle
\begin{abstract}
Within the kinetic energy driven superconducting mechanism, the
interplay between the single particle coherence and
superconducting instability in doped cuprates is studied. The
superconducting transition temperature increases with increasing
doping in the underdoped regime, and reaches a maximum in the
optimal doping, then decreases in the overdoped regime, however,
the values of this superconducting transition temperature in the
whole superconducting range are suppressed to low temperature due
to the single particle coherence. Within this superconducting
mechanism, we calculate the dynamical spin structure factor of
cuprate superconductors, and reproduce all main features of
inelastic neutron scattering experiments in the
superconducting-state.
\end{abstract}
]
\bigskip

\narrowtext

After intensive investigations over more than a decade, it has
become clear that the strong electron correlation in doped
cuprates plays a crucial role \cite{anderson1,hirsch}. The parent
compound of cuprates superconductors is a Mott insulator with the
antiferromagnetic (AF) long-range order (AFLRO), then changing the
carrier concentration by ionic substitution or increasing the
oxygen content turns these compounds into the superconducting
(SC)-state leaving the AF short-range correlation (AFSRC) still
intact \cite{kastner}. As a function of the hole doping
concentration, the SC transition temperature reaches a maximum in
the optimal doping, then decreases in both underdoped and
overdoped regimes \cite{tallon}. By virtue of systematic studies
using NMR and muon spin rotation techniques, particularly
inelastic neutron scattering, it has been well established that
AFSRC coexists with the SC-state in the whole SC regime
\cite{kastner,dai,yamada}, which implies a directive cooperative
relation between AFSRC and superconductivity.

In the conventional metals, superconductivity results when
electrons pair up into Cooper pairs, which is mediated by the
interaction of electrons with phonons \cite{bcs}. It has been
realized that this reduction in electron potential energy actually
corresponds to a decrease in the ionic kinetic energy, thus
providing a clear link between the pairing mechanism and phonons
\cite{chester}. As a natural consequence of this phonon-mediated
pairing, the Cooper pairs in the conventional superconductors have
an isotropic s-wave symmetry \cite{bcs}. In doped cuprates, the
charge carriers form the Cooper pairs when they become
superconductors as in the conventional superconductors
\cite{tsuei}. Although the possible doping dependent pairing
symmetry has been suggested \cite{yeh}, the Cooper pairs in the
optimally doped cuprate superconductors have a dominated d-wave
symmetry \cite{tsuei,yeh}, which is an indication of the
unconventional SC mechanism \cite{anderson1}. An alternative idea
is that superconductivity in doped cuprates arises directly from
the repulsive interactions between charge carriers
\cite{anderson2,anderson3}. In particular, it has been suggested
based on the non-Fermi liquid normal-state that the form of the
Cooper pairs is determined by the need to reduce the frustrated
kinetic energy \cite{hirsch,anderson3}. The normal-state exhibits
a number of anomalous properties which is due to the charge-spin
separation (CSS) \cite{anderson1,anderson2,anderson3}, while the
SC-state is characterized by the charge-spin recombination.

Recently, the angle resolved photoemission spectroscopy (ARPES)
measurements \cite{ding} have shown that the SC transition
temperature in doped cuprate is dependence of both gap parameter
and weight of the coherent excitations in the spectral function.
This strongly suggests that the single particle coherence plays an
important role in superconductivity. Within the $t$-$J$ model, one
of us \cite{feng1} has discussed the kinetic energy driven SC
mechanism in doped cuprates based on the CSS fermion-spin theory
\cite{feng2}, where the dressed holons interact occurring directly
through the kinetic energy by exchanging dressed spinon
excitations, leading to a net attractive force between dressed
holons, then the electron Cooper pairs originating from the
dressed holon pairing state are due to the charge-spin
recombination, and their condensation reveals the SC ground-state.
The SC transition temperature is identical to the dressed holon
pair transition temperature, and is proportional to the hole
doping concentration in the underdoped regime. However, the single
particle coherence in the system is not considered, which leads to
an obvious weakness that the SC transition temperature is too
high, and not suppressed in the overdoped regime. In this paper,
we cure this weakness by considering the charge carrier single
particle coherence, then we calculate explicitly the dynamical
spin structure factor (DSSF) of cuprate superconductors in terms
of the collective mode in the dressed holon particle-particle
channel, and reproduce all main features found in inelastic
neutron scattering experiments in the SC-state \cite{dai}.

We start from the $t$-$J$ model on a square lattice,
$H=-t\sum_{i\hat{\eta}\sigma}C^{\dagger}_{i\sigma}
C_{i+\hat{\eta}\sigma}+\mu \sum_{i\sigma}
C^{\dagger}_{i\sigma}C_{i\sigma}+J\sum_{i\hat{\eta}}{\bf S}_{i}
\cdot {\bf S}_{i+\hat{\eta}}$, with
$\hat{\eta}=\pm\hat{x},\pm\hat{y}$, $C^{\dagger}_{i\sigma}$
($C_{i\sigma}$) is the electron creation (annihilation) operator,
${\bf S}_{i}=C^{\dagger}_{i}{\vec\sigma} C_{i}/2$ is spin operator
with ${\vec\sigma}=(\sigma_{x}, \sigma_{y},\sigma_{z})$ as Pauli
matrices, and $\mu$ is the chemical potential. The $t$-$J$ model
is subject to an important on-site local constraint to avoid the
double occupancy, i.e., $\sum_{\sigma}C^{\dagger}_{i\sigma}
C_{i\sigma} \leq 1$. This local constraint can be treated properly
in analytical calculations within the CSS fermion-spin theory
\cite{feng2}, where the constrained electron operators are
decoupled as, $C_{i\uparrow}= h^{\dagger}_{i\uparrow} S^{-}_{i}$,
$C_{i\downarrow}= h^{\dagger}_{i\downarrow}S^{+}_{i}$, with the
spinful fermion operator $h_{i\sigma}= e^{-i\Phi_{i\sigma}}h_{i}$
describes the charge degree of freedom together with the phase
part of the spin degree of freedom (dressed holon), while the spin
operator $S_{i}$ describes the amplitude part of the spin degree
of freedom (dressed spinon), then the electron local constraint
for the single occupancy, $\sum_{\sigma}C^{\dagger}_{i\sigma}
C_{i\sigma}= S^{+}_{i}h_{i\uparrow}h^{\dagger}_{i\uparrow}
S^{-}_{i}+ S^{-}_{i} h_{i\downarrow}h^{\dagger}_{i\downarrow}
S^{+}_{i}=h_{i} h^{\dagger}_{i}(S^{+}_{i} S^{-}_{i}+S^{-}_{i}
S^{+}_{i})=1-h^{\dagger}_{i}h_{i}\leq 1$, is satisfied in
analytical calculations, and the double spinful fermion occupancy,
$h^{\dagger}_{i\sigma}h^{\dagger}_{i-\sigma}=e^{i\Phi_{i\sigma}}
h^{\dagger}_{i} h^{\dagger}_{i}e^{i\Phi_{i-\sigma}}=0$,
$h_{i\sigma}h_{i-\sigma}=e^{-i\Phi_{i\sigma}}h_{i}h_{i}
e^{-i\Phi_{i-\sigma}}=0$, are ruled out automatically. It has been
shown that these dressed holon and spinon are gauge invariant
\cite{feng2}, and in this sense, they are real \cite{laughlin}. At
the half-filling, the $t$-$J$ model is reduced to an AF Heisenberg
model, where there is no the charge degree of freedom, and the
real spinon excitation is described by the spin operator $S_{i}$.
Although in common sense $h_{i\sigma}$ is not a real spinful
fermion, it behaves like a spinful fermion. In this CSS
fermion-spin representation, the low-energy behavior of the
$t$-$J$ model can be expressed as \cite{feng2},
\begin{eqnarray}
H&=&-t\sum_{i\hat{\eta}}(h_{i\uparrow}S^{+}_{i}
h^{\dagger}_{i+\hat{\eta}\uparrow}S^{-}_{i+\hat{\eta}}+
h_{i\downarrow}S^{-}_{i}h^{\dagger}_{i+\hat{\eta}\downarrow}
S^{+}_{i+\hat{\eta}}) \nonumber \\
&-&\mu\sum_{i\sigma}h^{\dagger}_{i\sigma} h_{i\sigma}+J_{{\rm
eff}}\sum_{i\hat{\eta}}{\bf S}_{i}\cdot {\bf S}_{i+\hat{\eta}},
\end{eqnarray}
with $J_{{\rm eff}}=(1-x)^{2}J$, and $x=\langle
h^{\dagger}_{i\sigma}h_{i\sigma}\rangle=\langle h^{\dagger}_{i}
h_{i}\rangle$ is the hole doping concentration. The order
parameter for the electron Cooper pair in the CSS fermion-spin
approach can be expressed as, $\Delta=\langle
C^{\dagger}_{i\uparrow}C^{\dagger}_{j\downarrow}-
C^{\dagger}_{i\downarrow}C^{\dagger}_{j\uparrow}\rangle=\langle
h_{i\uparrow}h_{j\downarrow}S^{+}_{i}S^{-}_{j}-h_{i\downarrow}
h_{j\uparrow}S^{-}_{i}S^{+}_{j}\rangle$. In the doped regime
without AFLRO, the dressed spinons form the disordered spin liquid
state, where the dressed spinon correlation function $\langle
S^{+}_{i}S^{-}_{j}\rangle=\langle S^{-}_{i} S^{+}_{j}\rangle$,
then the order parameter for the electron Cooper pair can be
written as $\Delta=-\langle S^{+}_{i}S^{-}_{j} \rangle\Delta_{h}$,
with the dressed holon pairing order parameter $\Delta_{h}=\langle
h_{j\downarrow}h_{i\uparrow}-h_{j\uparrow}h_{i\downarrow}\rangle$.
This shows that the SC order parameter is closely related to the
dressed holon pairing amplitude, and is proportional to the number
of doped holes, and not to the number of electrons. However, in
the extreme low doped regime with AFLRO, where the dressed spinon
correlation function $\langle S^{+}_{i}S^{-}_{j}\rangle\neq\langle
S^{-}_{i} S^{+}_{j}\rangle$, then the conduct is disrupted by
AFLRO. Therefore in this paper, we only focus on the case without
AFLRO.

As shown in Ref. \cite{feng1}, the dressed holon-spinon coupling
occurring in the kinetic energy term of the $t$-$J$ model is quite
strong. This interaction can induce the dressed holon pairing
state by exchanging dressed spinon excitations in the higher power
of the hole doping concentration $x$. In this case, the SC
mechanism can be discussed in terms of Eliashberg's strong
coupling theory \cite{eliashberg}, and the self-consistent
equations that satisfied by the full dressed holon diagonal and
off-diagonal Green's functions are obtained as \cite{feng1},
\begin{mathletters}
\begin{eqnarray}
g(k)&=&g^{(0)}(k)\nonumber \\
&+&g^{(0)}(k)[\Sigma^{(h)}_{1}(k)g(k)-\Sigma^{(h)}_{2}(-k)
\Im^{\dagger}(k)], \\
\Im^{\dagger}(k)&=&g^{(0)}(-k)[\Sigma^{(h)}_{1}(-k)
\Im^{\dagger}(-k)+\Sigma^{(h)}_{2}(-k)g(k)],
\end{eqnarray}
\end{mathletters}
respectively, where the four-vector notation $k=({\bf k},
i\omega_{n})$, the dressed holon mean-field (MF) diagonal Green's
function \cite{feng1,feng2} $g^{(0)-1}(k)=i\omega_{n}-\xi_{{\bf k}
}$, the MF dressed holon excitation spectrum $\xi_{{\bf k}}=
Zt\chi \gamma_{{\bf k}}-\mu$, with $\gamma_{{\bf k}}=(1/Z)
\sum_{\hat{\eta}}e^{i{\bf k}\cdot \hat{\eta}}$, $Z$ is the number
of the nearest neighbor sites, the dressed spinon correlation
function $\chi=\langle S_{i}^{+}S_{i+\hat{\eta}}^{-}\rangle$, and
the dressed holon self-energies \cite{feng1},
\begin{mathletters}
\begin{eqnarray}
\Sigma^{(h)}_{1}(k)&=&(Zt)^{2}{1\over N^{2}}\sum_{{\bf p,p'}}
\gamma^{2}_{{\bf p+p'+k}}{1\over \beta}\sum_{ip_{m}}g(p+k)
\nonumber \\
&\times&{1\over\beta}\sum_{ip'_{m}}D^{(0)}(p')D^{(0)}(p'+p), \\
\Sigma^{(h)}_{2}(k)&=&(Zt)^{2}{1\over N^{2}}\sum_{{\bf p,p'}}
\gamma^{2}_{{\bf p+p'+k}}{1\over \beta}\sum_{ip_{m}}\Im (-p-k)
\nonumber \\
&\times& {1\over\beta}\sum_{ip'_{m}}D^{(0)}(p')D^{(0)}(p'+p),
\end{eqnarray}
\end{mathletters}
where $p=({\bf p},ip_{m})$, $p'=({\bf p'},ip_{m}')$, the MF
dressed spinon Green's function \cite{feng1,feng2},
$D^{(0)-1}(p)=[ (ip_{m})^{2}-\omega_{{\bf p}}^{2}]/B_{{\bf p}}$,
with $B_{{\bf p}} =\lambda[2\chi_{z}(\epsilon\gamma_{{\bf
p}}-1)+\chi(\gamma_{{\bf p}}-\epsilon)]$, $\lambda=2ZJ_{{\rm
eff}}$, $\epsilon=1+2t\phi/ J_{{\rm eff}}$, and the MF dressed
spinon excitation spectrum $\omega^{2}_{{\bf
p}}=A_{1}\gamma^{2}_{{\bf p}}+A_{2}\gamma_{{\bf p}}+A_{3}$, with
$A_{1}=\alpha\epsilon\lambda^{2}(\epsilon \chi_{z}+\chi/2)$,
$A_{2}=-\epsilon\lambda^{2}[\alpha (\chi_{z}+
\epsilon\chi/2)+(\alpha C_{z}+(1-\alpha)/(4Z)-\alpha\epsilon\chi
/(2Z))+(\alpha C+(1-\alpha)/(2Z)-\alpha\chi_{z}/2)/2]$, $A_{3} =
\lambda^{2}[\alpha C_{z}+(1-\alpha)/(4Z)-\alpha\epsilon\chi /(2Z)+
\epsilon^{2}(\alpha C+(1-\alpha)/(2Z)-\alpha \chi_{z}/2) /2]$, and
the dressed holon particle-hole parameter $\phi=\langle
h^{\dagger}_{i\sigma}h_{i+\hat{\eta}\sigma}\rangle$, the dressed
spinon correlation functions $\chi_{z}=\langle S_{i}^{z}
S_{i+\hat{\eta}}^{z}\rangle$, $C=(1/Z^{2})
\sum_{\hat{\eta},\hat{\eta'}}\langle S_{i+\hat{\eta}}^{+}
S_{i+\hat{\eta'}}^{-}\rangle$, $C_{z}=(1/Z^{2})
\sum_{\hat{\eta},\hat{\eta'}}\langle S_{i+\hat{\eta}}^{z}
S_{i+\hat{\eta'}}^{z}\rangle$. In order to satisfy the sum rule of
the dressed spinon correlation function $\langle S^{+}_{i}
S^{-}_{i}\rangle=1/2$ in the case without AFLRO, the important
decoupling parameter $\alpha$ has been introduced in the MF
calculation \cite{feng3}, which can be regarded as the vertex
correction. In the above calculations of the self-energies
\cite{feng1}, the dressed spinon part has been limited to the MF
level, i.e., the full dressed spinon Green's function $D(p)$ in
Eq. (3) has been replaced by the MF dresed spinon Green's
function, since the normal-state charge transport obtained at this
level can well describe the experimental data \cite{feng2}.

Since the pairing force and dressed holon gap function have been
incorporated into the self-energy function $\Sigma^{(h)}_{2}(k)$,
then it is called as the effective dressed holon gap function. On
the other hand, the self-energy function $\Sigma^{(h)}_{1}(k)$
renormalizes the MF dressed holon spectrum, and therefore it
describes the dressed holon single particle coherence. In other
words, $\Sigma^{(h)}_{1}(k)$ describes the dressed holon quantum
fluctuation, and $\Sigma^{(h)}_{2}(k)$ describes the dressed holon
pairing instability. Moreover, $\Sigma^{(h)}_{2}(k)$ is an even
function of $i\omega_{n}$, while $\Sigma^{(h)}_{1}(k)$ is not. In
this case, it is convenient to break $\Sigma^{(h)}_{1}(k)$ up into
its symmetric and antisymmetric parts as,
$\Sigma^{(h)}_{1}(k)=\Sigma^{(h)}_{1e}(k)+i\omega_{n}
\Sigma^{(h)}_{1o}(k)$, where $\Sigma^{(h)}_{1e}(k)$ and
$\Sigma^{(h)}_{1o}(k)$ are both even functions of $i\omega_{n}$.
Now we define the dressed holon renormalization coefficient
(charge carrier weight of the coherent excitations in the spectral
function) $Z_{F}(k)=1- \Sigma^{(h)}_{1o}(k)$. As in the
conventional superconductor \cite{eliashberg}, the retarded
function ${\rm Re}\Sigma^{(h)}_{1e} (k)$ may be a constant,
independent of (${\bf k},\omega$). It just renormalizes the
chemical potential, and therefore can be neglected. Furthermore,
we only study the static limit of the effective dressed holon gap
function and dressed holon renormalization coefficient, i.e.,
$\Sigma^{(h)}_{2}(k)= \bar{\Delta}_{h}({\bf k})$, and $Z_{F}({\bf
k})=1- \Sigma^{(h)}_{1o}({\bf k})$. In this case, the dressed
holon diagonal and off-diagonal Green's functions in Eq. (2) can
be rewritten explicitly as,
\begin{mathletters}
\begin{eqnarray}
g(k)&=&{1\over 2Z_{F}({\bf k})}\left (1+  {\bar{\xi_{{\bf k}}}
\over E_{{\bf k}}} \right ){1\over i\omega_{n}-E_{{\bf k}}}
\nonumber \\
&+&{1\over 2Z_{F}({\bf k})}\left (1- {\bar{\xi_{{\bf k}}}\over
E_{{\bf k}}}\right ){1\over i\omega_{n}+E_{{\bf k}}}, \\
\Im^{\dagger}(k)&=&-{1\over Z_{F}({\bf k})}{\bar{\Delta}_{hZ}({\bf
k})\over 2E_{{\bf k}}}\left ( {1\over i\omega_{n}-E_{{\bf k}}}-
{1\over i\omega_{n}+ E_{{\bf k}}}\right ),
\end{eqnarray}
\end{mathletters}
with $\bar{\xi_{{\bf k}}}=\xi_{{\bf k}}/Z_{F}({\bf k})$,
$\bar{\Delta}_{hZ}({\bf k})=\bar{\Delta}_{h}({\bf k})/Z_{F}({\bf
k})$, and the dressed holon quasiparticle spectrum $E_{{\bf k}}=
\sqrt{\bar{\xi^{2}_{{\bf k}}}+\mid\bar{\Delta}_{hZ}({\bf k})
\mid^{2}}$. Although $Z_{F}({\bf k})$ is still a function of ${\bf
k}$, the wave vector dependence is unimportant, since everything
happens at the electron Fermi surface. Therefore we need to
estimate the special wave vector ${\bf k}_{0}$ that guarantees
$Z_{F}=Z_{F} ({\bf k}_{0})$ near the electron Fermi surface. In
the present CSS fermion-spin framework \cite{feng2}, the electron
diagonal Green's function $G(i-j,t-t')=\langle\langle C_{i\sigma}
(t);C^{\dagger}_{j\sigma}(t')\rangle\rangle$ is a convolution of
the dressed spinon Green's function $D(p)$ and dressed holon
diagonal Green's function $g(k)$, which reflects the charge-spin
recombination \cite{anderson3}, and can be calculated as
\cite{feng3},
\begin{eqnarray}
G(k)&=&{1\over N}\sum_{{\bf p}}\int^{\infty}_{-\infty}{d\omega'
\over 2\pi}{d\omega''\over 2\pi}A_{s}({\bf p},\omega') A_{h}({\bf
p-k},\omega'') \nonumber \\
&\times& {n_{F}(\omega'')+n_{B}(\omega')\over i\omega_{n}
+\omega''-\omega'},
\end{eqnarray}
where the dressed spinon spectral function $A_{s}({\bf k},\omega)
=-2{\rm Im}D({\bf k},\omega)$, the dressed holon spectral function
$A_{h}({\bf k},\omega) =-2{\rm Im}g({\bf k},\omega)$, and $n_{B}
(\omega)$ and $n_{F}(\omega)$ are the boson and fermion
distribution functions, respectively. This electron diagonal
Green's function has been used to extract the electron momentum
distribution (then the electron Fermi surface) as \cite{feng3},
\begin{eqnarray}
n_{{\bf k}}={1\over 2}-{1\over N}\sum_{{\bf p}}n_{s}({\bf p})
\int^{\infty}_{-\infty}{d\omega\over 2\pi}A_{h}({\bf p-k},
\omega)n_{F}(\omega),
\end{eqnarray}
with $n_{s}({\bf p})=\int^{\infty}_{-\infty}d\omega A_{s}({\bf p},
\omega)n_{s}(\omega)/2\pi$ is the dressed spinon momentum
distribution. In the present case, this electron momentum
distribution can be evaluated in terms of the MF dressed spinon
Green's function and dressed holon diagonal Green's function (4a)
as,
\begin{eqnarray}
n_{{\bf k}}={1\over 2}&-&{1\over N}\sum_{{\bf p}}n^{(0)}_{s}({\bf
p}){1\over 2Z_{F}({\bf p-k})} \nonumber \\
&\times& \left (1- {\bar{\xi}_{{\bf p-k}}\over E_{{\bf p-k}}}{\rm
tanh} [{1\over 2}\beta E_{{\bf p-k}}]\right ),
\end{eqnarray}
with $n^{(0)}_{s}({\bf p})=B_{{\bf p}}{\rm coth}(\beta\omega_{{\bf
p}}/2)/(2\omega_{{\bf p}})$. Since the dressed spinons center
around $[\pm\pi,\pm\pi]$ in the Brillouin zone in the MF level
\cite{feng3}, then the electron momentum distribution (7) can be
approximately reduced as $n_{{\bf k}}\approx 1/2-\rho^{(0)}_{s}[1-
\bar{\xi}_{{\bf k_{A}-k}}{\rm tanh}(\beta E_{{\bf k_{A}-k}}/2)/
E_{{\bf k_{A}-k}}]/(2Z_{F})$, with ${\bf k_{A}}=[\pi,\pi]$, and
$\rho^{(0)}_{s}=(1/N)\sum_{{\bf p}=(\pm\pi,\pm\pi)}n^{(0)}_{s}
({\bf p})$. Therefore the Fermi wave vector from this electron
momentum distribution is estimated \cite{feng3} as ${\bf k_{F}}
\approx [(1-x)\pi/2,(1-x)\pi/2]$, which evolves with doping. In
this case, the wave vector ${\bf k}_{0}$ is obtained as ${\bf
k}_{0}={\bf k_{A}} -{\bf k_{F}}$, then we only need to calculate
$Z_{F}=Z_{F}({\bf k}_{0})$ as mentioned above. Since the
charge-spin recombination from the convolution of the dressed
spinon Green's function and dressed holon diagonal Green's
function leads to form the electron Fermi surface
\cite{anderson3}, then the dressed holon single particle coherence
$Z_{F}$ appearing in the electron momentum distribution also
reflects the electron single particle coherence.

ARPES measurements \cite{shen1} have shown that in the real space
the gap function and pairing force have a range of one lattice
spacing, this indicates that the effective dressed holon gap
function can be expressed as $\bar{\Delta}_{hZ}({\bf k})
=\bar{\Delta}^{(a)}_{hZ}\gamma^{(a)}_{{\bf k}}$. On the other
hand, some experiments seem consistent with an s-wave pairing
\cite{chaudhari}, while other measurements gave the evidence in
favor of the d-wave pairing \cite{martindale,tsuei}, therefore in
the following discussions, we consider the cases of
$\bar{\Delta}^{(a)}_{hZ}=\bar{\Delta}^{(s)}_{hZ}$, with
$\gamma^{(a)}_{{\bf k}}=\gamma^{(s)}_{{\bf k}}=\gamma_{{\bf k}}=
({\rm cos}k_{x}+{\rm cos}k_{y})/2$, for the s-wave pairing, and
$\bar{\Delta}^{(a)}_{hZ}=\bar{\Delta}^{(d)}_{hZ}$,
$\gamma^{(a)}_{{\bf k}}=\gamma^{(d)}_{{\bf k}}= ({\rm cos}k_{x}-
{\rm cos}k_{y})/2$, for the d-wave pairing, respectively. In this
case, the dressed holon effective gap parameter and
renormalization coefficient in Eq. (3) satisfy the following
equations \cite{feng1},
\begin{mathletters}
\begin{eqnarray}
1&=&(Zt)^{2}{1\over N^{3}}\sum_{{\bf k,q,p}}\gamma^{2}_{{\bf k+q}}
\gamma^{(a)}_{{\bf k-p+q}}\gamma^{(a)}_{{\bf k}}{1\over Z^{2}_{F}
E_{{\bf k}}}{B_{{\bf q}}B_{{\bf p}}\over\omega_{{\bf q}}
\omega_{{\bf p}}} \nonumber \\
&\times&\left({F^{(1)}_{1}({\bf k,q,p})\over (\omega_{{\bf p}}-
\omega_{{\bf q}})^{2}-E^{2}_{{\bf k}}}+{F^{(2)}_{1}({\bf k,q,p})
\over (\omega_{{\bf p}}+\omega_{{\bf q}})^{2}- E^{2}_{{\bf k}}}
\right ) ,\\
Z_{F}=1&+&(Zt)^{2}{1\over N^{2}}\sum_{{\bf q,p}}\gamma^{2}_{{\bf
p+k_{0}}}{1\over Z_{F}}{B_{{\bf q}}B_{{\bf p}}\over 4\omega_{{\bf
q}}\omega_{{\bf p}}} \nonumber \\
&\times& \left({F^{(1)}_{2}({\bf q,p})\over (\omega_{{\bf p}}-
\omega_{{\bf q}}-E_{{\bf p-q+k_{0}}})^{2}}\right . \nonumber \\
&+&{F^{(2)}_{2}({\bf q,p}) \over (\omega_{{\bf p}}-\omega_{{\bf
q}}+E_{{\bf p-q+k_{0}}})^{2}} \nonumber \\
&+& {F^{(3)}_{2}({\bf q,p})\over (\omega_{{\bf p}}+ \omega_{{\bf
q}}-E_{{\bf p-q+k_{0}}})^{2}} \nonumber \\
&+& \left . {F^{(4)}_{2}({\bf q,p})\over (\omega_{{\bf
p}}+\omega_{{\bf q}}+E_{{\bf p-q+k_{0}}})^{2}} \right ) ,
\end{eqnarray}
\end{mathletters}
respectively, where $F^{(1)}_{1}({\bf k,q,p})=(\omega_{{\bf p}}-
\omega_{{\bf q}})[n_{B}(\omega_{{\bf q}})-n_{B}(\omega_{{\bf p}})]
[1-2 n_{F}(E_{{\bf k}})]+E_{{\bf k}}[n_{B}(\omega_{{\bf p}})n_{B}(
-\omega_{{\bf q}})+n_{B}(\omega_{{\bf q}})n_{B}(-\omega_{{\bf p}})
]$, $F^{(2)}_{1}({\bf k,q,p})=-(\omega_{{\bf p }}+\omega_{{\bf
q}}) [n_{B}(\omega_{{\bf q}})-n_{B}(-\omega_{{\bf p}})][1-2 n_{F}
(E_{{\bf k}})]+E_{{\bf k}}[n_{B}(\omega_{{\bf p}}) n_{B}
(\omega_{{\bf q}})+n_{B}(-\omega_{{\bf p}})n_{B}(-\omega_{{\bf q}
})]$, $F^{(1)}_{2}({\bf q,p})=n_{F}(E_{{\bf p- q+k_{0}}})[n_{B}
(\omega_{{\bf q}})-n_{B}(\omega_{{\bf p}})]- n_{B}(\omega_{{\bf
p}})n_{B}(-\omega_{{\bf q}})$, $F^{(2)}_{2} ({\bf q,p})=
n_{F}(E_{{\bf p-q+k_{0}}}) [n_{B}(\omega_{{\bf p}})-n_{B}
(\omega_{{\bf q}})]-n_{B}(\omega_{{\bf q}})n_{B} (-\omega_{{\bf
p}})$, $F^{(3)}_{2}({\bf q,p})= n_{F}(E_{{\bf p-q+k_{0}}})
[n_{B}(\omega_{{\bf q}})-n_{B}(-\omega_{{\bf p}})]+n_{B}
(\omega_{{\bf p}})n_{B}(\omega_{{\bf q}})$, and $F^{(4)}_{2}({\bf
q,p})=n_{F}(E_{{\bf p-q+k_{0}}})[n_{B}(-\omega_{{\bf q}})-n_{B}
(\omega_{{\bf p}})]+n_{B}(-\omega_{{\bf p}})n_{B}(-\omega_{{\bf
q}})$. These two equations must be solved simultaneously with
other self-consistent equations \cite{feng1}, then all order
parameters, decoupling parameter $\alpha$, and chemical potential
$\mu$ are determined by the self-consistent calculation
\cite{feng3}. In this cas, the dressed holon pair order parameter
is obtained in terms of the off-diagonal Green's function (4b) as,
$\Delta^{(a)}_{h}=(2/N)\sum_{{\bf k}}[\gamma^{(a)}_{{\bf k}} ]^{2}
\bar{\Delta}^{(a)}_{hZ}{\rm tanh}[\beta E_{{\bf k}}/2]/
(Z_{F}E_{{\bf k}})$. As shown in Ref. \cite{feng1}, the dressed
holon pairing state originating from the kinetic energy term by
exchanging dressed spinon excitations also lead to form the
electron Cooper pairing state, and the SC gap function is obtained
from the electron off-diagonal Green's function
$\Gamma^{\dagger}(i-j,t-t') =\langle\langle
C^{\dagger}_{i\uparrow}(t);C^{\dagger}_{j\downarrow}(t')\rangle
\rangle$, which is a convolution of the dressed spinon Green's
function and dressed holon off-diagonal Green's function
\cite{anderson3}, and has been obtained \cite{feng1} in terms of
the MF dressed spinon Green's function and dressed holon
off-diagonal Green's function (4b), then the SC gap function is
obtained from this electron off-diagonal Green's function as,
\begin{eqnarray}
\Delta^{(a)}({\bf k})&=&-{1\over N}\sum_{{\bf p}}
{\bar{\Delta}^{(a)}_{Zh}({\bf p-k})\over 2Z_{F}E_{{\bf p-k}}}{\rm
tanh}[{1\over 2}\beta E_{{\bf p-k}}]\nonumber \\
&\times& {B_{{\bf p}}\over 2\omega_{{\bf p}}} {\rm coth}[{1\over
2}\beta\omega_{{\bf p}}],
\end{eqnarray}
this shows that the symmetry of the electron Cooper pair is
determined by the symmetry of the dressed holon pair, and
therefore the SC gap function can be written as $\Delta^{(a)}({\bf
k})=\Delta^{(a)}\gamma^{(a)}_{{\bf k}}$, with the SC gap parameter
is evaluated \cite{feng1} in terms of the dressed holon pair order
parameter and Eq. (9) as $\Delta^{(a)}=-\chi\Delta^{(a)}_{h}$. The
present result in Eq. (9) also shows that the SC transition
temperature $T^{(a)}_{c}$ occurring in the case of the SC gap
parameter $\Delta^{(a)}=0$ is identical to the dressed holon pair
transition temperature occurring in the case of the effective
dressed holon pairing gap parameter $\bar{\Delta}^{(a)}_{hZ}=0$.
The SC transition temperature $T^{(a)}_{c}$ as a function of the
hole doping concentration $x$ in the s-wave symmetry (solid line)
and d-wave symmetry (dashed line) for $t/J=2.5$ is plotted in Fig.
1 in comparison with the experimental result \cite{tallon}
(inset). For the s-wave symmetry, the maximal SC transition
temperature T$^{(s)}_{c}$ occurs around a particular doping
concentration $x\approx 0.11$, and then decreases for both lower
doped and higher doped regimes. However, for the d-wave symmetry,
the maximal SC transition temperature T$^{(d)}_{c}$ occurs around
the optimal doping concentration $x\approx 0.18$, and then
decreases for both underdoped and overdoped regimes. Although the
SC pairing symmetry is doping dependent, the SC state has the
d-wave symmetry in a wide range of doping, in qualitative
agreement with the experiments \cite{yeh,biswas}. Furthermore,
T$^{(d)}_{c}$ in the underdoped regime (T$^{(s)}_{c}$ in the lower
doped regime) is proportional to the hole doping concentration
$x$, and therefore T$^{(d)}_{c}$ in the underdoped regime
(T$^{(s)}_{c}$ in the lower doped regime) is set by the hole
doping concentration, this reflects that the dressed holon density
directly determines the superfluid density in the underdoped
regime for the d-wave case (the lower doped regime for the s-wave
case). Using an reasonable estimation value of $J\sim 800$K to
1200K in doped cuprates, the SC transition temperature in the
optimal doping is T$^{(d)}_{c} \approx 0.2J\approx 160{\rm K}\sim
240{\rm K}$, also in qualitative agreement with the experimental
data \cite{tallon,biswas}.

\begin{figure}[prb]
\epsfxsize=3.5in\centerline{\epsffile{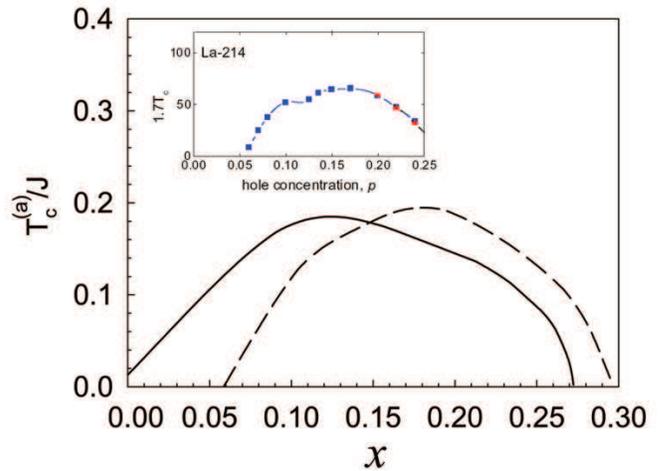}}\caption{The
superconducting transition temperature as a function of the hole
doping concentration in the s-wave symmetry (solid line) and
d-wave symmetry (dashed line) for $t/J=2.5$. Inset: the
experimental result taken from Ref. [4]. }
\end{figure}

In the framework of the kinetic energy driven superconductivity
\cite{feng1}, $\Sigma^{(h)}_{1o}({\bf k})$ (then $Z_{F}$)
describes the single particle coherence, which favors the single
dressed holon motion in the background of the dressed spinon
fluctuation, while $\Sigma^{(h)}_{2}({\bf k})$ describes the
effective dressed holon pairing gap parameter, which measures the
strength of the binding of dressed holons into dressed holon pairs
and favors the dressed holon pair motion, therefore there is a
competition between the single particle coherence and SC
instability. In the underdoped and optimally doped regimes, both
superfluid density and $Z_{F}$ increase with increasing doping,
this leads to that the SC transition temperature increases with
increasing doping, and is proportional to the hole doping
concentration \cite{feng1}. In the overdoped regime, although the
superfluid density still increases with increasing doping
\cite{feng1}, $Z_{F}$ is slows down with increasing doping
\cite{ding}, which leads to that the SC transition temperature
decreases with increasing doping in the overdoped regime. However,
as a result of the competition and self-consistent motion of the
dressed holons, dressed holon pairs, and dressed spinons in the
whole SC regime, the SC transition temperature is suppressed to
the lower temperature due to the single particle coherence, this
is why the SC transition temperature is so low in doped cuprates.

Now we turn to discuss the convergence of energy dependent
incommensurate (IC) scattering to commensurate resonance, which is
one of the most striking features of cuprate superconductors
\cite{anderson1,hirsch,kastner}. Experimentally NMR and inelastic
neutron scattering have provided rather detailed information on
the spin fluctuation \cite{kastner,dai,yamada}, where the distinct
phenomena are the presence of the IC scattering peaks at low
energies and commensurate resonance peak at relatively high
energies, i.e., the IC scattering peaks are shifted from the AF
wave vector [$\pi$,$\pi$] to four points [$\pi (1\pm\delta),\pi$]
and [$\pi,(1\pm \delta)\pi$] (in units of inverse lattice
constant) at low energies with $\delta $ as the incommensurability
parameter, which depends on both hole doping concentration and
energy, then a sharp resonance peak at the commensurate AF wave
vector [$\pi$,$\pi$] is observed at relatively high energies.
Although some of these magnetic properties have been observed in
doped cuprates in the normal-state, these IC scattering and
commensurate resonance are the main new feature that appears into
the SC-state \cite{kastner,dai}.

Within the CSS fermion-spin theory, the IC scattering and
integrated spin response in the {\it normal-state} have been
discussed \cite{feng2}, and the results of the doping dependence
of the IC parameter $\delta$ and integrated dynamical spin
susceptibility are consistent with experiments in the {\it
normal-state} \cite{kastner,yamada}. Since the AF fluctuation is
dominated by the scattering of the dressed spinons in the CSS
fermion-spin theory \cite{feng2}, while in the present case in the
SC state, this AF fluctuation has been incorporated into the
electron off-diagonal Green's function (hence the electron Cooper
pair) in terms of the dressed spinon Green's function, therefore
there is a coexistence of the electron Cooper pair and AFSRC, and
then AFSRC can persist into superconductivity \cite{feng1}.
Following the previous discussions for the normal-state case
\cite{feng2}, DSSF in the SC-state with the d-wave symmetry can be
obtained as,
\begin{eqnarray}
&S&({\bf k},\omega)=-2[1+n_{B}(\omega)]{\rm Im}D({\bf k},\omega)=
2[1+n_{B}(\omega)]\nonumber \\
&\times& {B^{2}_{{\bf k}}{\rm Im}\Sigma^{(s)}({\bf k},\omega)\over
[\omega^{2}-\omega^{2}_{{\bf k}}-B_{{\bf k}}{\rm
Re}\Sigma^{(s)}({\bf k},\omega)]^{2}+[B_{{\bf k}}{\rm Im}
\Sigma^{(s)}({\bf k},\omega)]^{2}},
\end{eqnarray}
where the full dressed spinon Green's function, $D^{-1}({\bf k},
\omega)= D^{(0)-1}({\bf k},\omega)-\Sigma^{(s)}({\bf k},\omega)$,
with ${\rm Im}\Sigma^{(s)}({\bf k},\omega)$ and ${\rm Re}
\Sigma^{(s)}({\bf k}, \omega)$ are the imaginary and real parts of
the second order spinon self-energy, respectively, obtained from
the dressed holon bubble in the dressed holon particle-particle
channel as,
\begin{eqnarray}
\Sigma^{(s)}({\bf k},\omega)&=&(Zt)^{2}{1\over N^{2}}\sum_{{\bf
p,q}}(\gamma^{2}_{{\bf q+p+k}}+\gamma^{2}_{{\bf p-k}}) \nonumber
\\
&\times& {B_{{\bf q+k}}\over\omega_{{\bf q+k}}}
{\bar{\Delta}^{(a)}_{hZ}({\bf p}) \bar{\Delta}^{(a)}_{hZ}({\bf
p+q})\over 4Z^{2}_{F}E_{{\bf p}}E_{{\bf p+q}}} \nonumber \\
&\times& \left ( {F^{(1)}_{s}({\bf k,p,q})\over \omega^{2}-
(E_{{\bf p}} -E_{{\bf p+q}}+\omega_{{\bf q+k}})^{2}} \right .
\nonumber \\
&+& {F^{(2)}_{s}({\bf k,p,q})\over \omega^{2}-(E_{{\bf p+q}}
-E_{{\bf p}}+\omega_{{\bf q+k}})^{2}} \nonumber\\
&+& {F^{(3)}_{s}({\bf k,p,q})\over \omega^{2}-(E_{{\bf p}} +
E_{{\bf p+q}} +\omega_{{\bf q+k}})^{2}} \nonumber \\
&+&\left . {F^{(4)}_{s}({\bf k,p,q})\over \omega^{2}- (E_{{\bf
p+q}}+E_{{\bf p}}-\omega_{{\bf q+k}})^{2}} \right ),
\end{eqnarray}
with $F^{(1)}_{s}({\bf k,p,q}) = (E_{{\bf p}} - E_{{\bf p+q}} +
\omega_{{\bf q+k}}) \{n_{B}(\omega_{{\bf q+k}}) [n_{F}(E_{{\bf
p}}) - n_{F}(E_{{\bf p+q}})] - n_{F}(E_{{\bf p+q}}) n_{F}(-E_{{\bf
p}})\}$, $F^{(2)}_{s}({\bf k,p,q}) = (E_{{\bf p+q}} - E_{{\bf p}}
+ \omega_{{\bf q+k}}) \{n_{B}(\omega_{{\bf q+k}}) [n_{F}(E_{{\bf
p+q}}) - n_{F}(E_{{\bf p}})] - n_{F}(E_{{\bf p}}) n_{F}(-E_{{\bf
p+q}})\}$, $F^{(3)}_{s}({\bf k,p,q}) = (E_{{\bf p}} + E_{{\bf
p+q}} + \omega_{{\bf q+k}})\{n_{B}(\omega_{{\bf q+k}})
[n_{F}(-E_{{\bf p}}) - n_{F}(E_{{\bf p+q}})] + n_{F}(-E_{{\bf
p+q}}) n_{F}(-E_{{\bf p}})\}$, $F^{(4)}_{s}({\bf k,p,q}) =
(E_{{\bf p}} + E_{{\bf p+q}} - \omega_{{\bf q+k}}) \{n_{B}
(\omega_{{\bf q+k}}) [n_{F}(-E_{{\bf p}}) - n_{F}(E_{{\bf p+q}})]
-n_{F}(E_{{\bf p+q}}) n_{F}(E_{{\bf p}})\}$.

\begin{figure}[prb]
\epsfxsize=3.5in\centerline{\epsffile{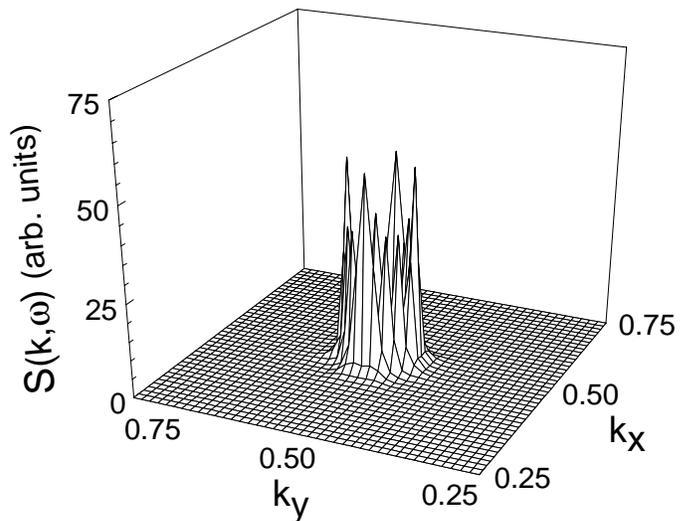}}\caption{The
dynamical spin structure factor $S({\bf k},\omega)$ in the
($k_{x},k_{y}$) plane in the superconducting-state at $x=0.15$ in
$T=0.002J$ and $\omega =0.11J$ for $t/J=2.5$.}
\end{figure}

\begin{figure}[prb]
\epsfxsize=3.5in\centerline{\epsffile{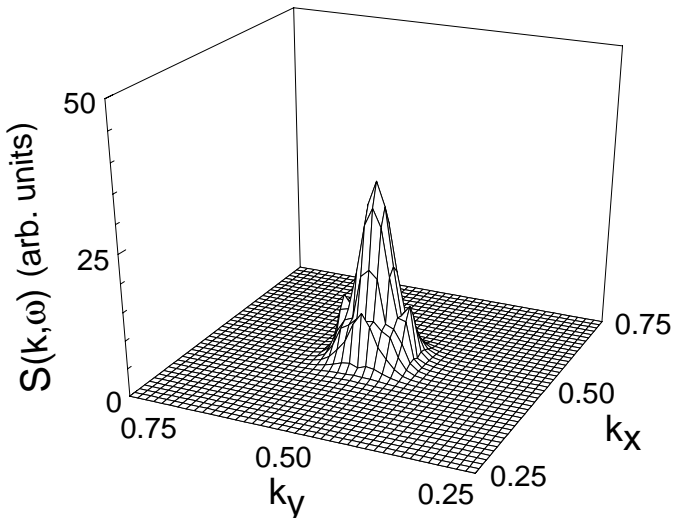}}\caption{The
dynamical spin structure factor $S({\bf k},\omega)$ in the
superconducting-state at $x=0.15$ for $t/J=2.5$ in $T=0.002J$ and
$\omega =0.33J$.}
\end{figure}

In Fig. 2, we plot $S({\bf k}, \omega)$ in the ($k_{x},k_{y}$)
plane at doping $x=0.15$ in temperature $T=0.002J$ and energy
$\omega =0.11J$ for parameter $t/J=2.5$, which shows that the IC
spin fluctuation pattern occurs with doping, and the IC peaks are
located at $[(1\pm\delta) /2,1/2]$ and $[1/2,(1\pm \delta )/2]$
(hereafter we use the units of $[2\pi ,2\pi ]$). For considering
the resonance at relatively high energies we have made a series of
scans for $S({\bf k},\omega )$ at different energies, and the
result at $x=0.15$ for $t/J=2.5$ in $T=0.002J$ and $\omega =0.33J$
is shown in Fig. 3. Comparing it with Fig. 2 for the same set of
parameters except for $\omega =0.33J$, we find that IC peaks are
energy dependent, i.e., although these scattering peaks retain the
IC pattern at $[(1\pm \delta )/2,1/2]$ and $[1/2,(1\pm \delta
)/2]$ in low energies, the positions of IC peaks move towards
$[1/2,1/2]$ with increasing energy, and then the commensurate
$[1/2,1/2]$ resonance peak appears at relatively high energies
$\omega_{r}=0.33J$. Moreover, the resonance energy is doping
dependent, and is proportional to $x$ in the underdoped regime
\cite{feng9}. Our these results are in qualitative agreement with
experiments of doped cuprates in the SC-state \cite{dai}.

As in the normal-state case \cite{feng2}, the physics of the
convergence of the IC magnetic scattering peaks at lower energies
to commensurate resonance at higher energies in the SC-state also
can be understood from the properties of the renormalized dressed
spinon excitation spectrum $\Omega^{2}_{{\bf k}}= \omega^{2}_{{\bf
k}}+ {\rm Re} \Sigma^{(s)}({\bf k},\Omega_{{\bf k}})$, which is
doping and energy dependent. DSSF in Eq. (10) has a well-defined
resonance character, where $S({\bf k},\omega)$ exhibits peaks when
the incoming neutron energy $\omega$ is equal to the renormalized
spin excitation (the collective mode in the dressed holon
particle-particle channel), i.e., $W({\bf k}_{c}, \omega)\equiv
[\omega^{2}-\omega_{{\bf k}_{c}}^{2}-B_{{\bf k}_{c}}{\rm Re}
\Sigma^{(s)}({\bf k}_{c}, \omega)]^{2}=[\omega^{2}- \Omega_{{\bf
k}_{c}}^{2}]^{2}\sim 0$ for certain critical wave vectors ${\bf
k}_{c}$, then the weight of these peaks is dominated by $1/{\rm
Im}\Sigma^{(s)}({\bf k}_{c},\omega)$. In this case, the positions
of the magnetic scattering peaks are determined by both functions
$W({\bf k},\omega)$ and ${\rm Im} \Sigma^{(s)}({\bf k}, \omega)$.
Within the kinetic energy driven superconductivity, as a result of
self-consistent motion of the dressed holon pairs and spinons, the
IC scattering is developed beyond certain critical doping at low
energies, this reflects that the low energy spin excitations drift
away from the AF wave vector, or the zero point of $W({\bf k}_{c},
\omega)$ is shifted from $[1/2,1/2]$ to ${\bf k}_{c}$. With
increasing energy, the spin excitations move towards to
$[1/2,1/2]$, i.e., the zero point of $W({\bf k}_{c},\omega)$ in
${\bf k}_{c}$ turns back to $[1/2,1/2]$, then the commensurate
$[1/2,1/2]$ resonance appears at relatively high resonance energy
$\omega_{r}$. Since the essential physics is dominated by the
dressed spinon self-energy renormalization due to the dressed
holon bubble in the dressed holon particle-particle channel, then
in this sense the mobile dressed holon pairs are the key factor
leading to the convergence of the IC scattering peaks at lower
energies to commensurate resonance at higher energies, i.e., the
mechanism of the IC scattering peaks and commensurate resonance in
the SC state is most likely related to the motion of the dressed
holon pairs. This is why the position of the IC magnetic
scattering peaks and commensurate resonance in the SC-state can be
determined in the present study within the $t$-$J$ model, while
the dressed spinon energy dependence is ascribed purely to the
self-energy effects which arise from the the dressed holon bubble
in the dressed holon particle-particle channel.

In summary, within the CSS fermion-spin theory, we have discussed
the interplay between the single particle coherence and kinetic
energy driven SC instability in doped cuprates. The dressed holon
pair instability is caused directly through the kinetic energy by
exchanging dressed spinon excitations, then the electron Cooper
pairs originating from the dressed holon pairing state are due to
the charge-spin recombination, and their condensation reveals the
SC ground-state \cite{feng1}. The SC transition temperature
$T_{c}$ is determined by the dressed holon pair transition
temperature, and is suppressed to low temperature due to the
single particle coherence. Although the symmetry of the SC-state
is doping dependent, the SC-state has the d-wave symmetry in a
wide range of doping. Moreover, the maximal SC transition
temperature $T^{(d)}_{c}$ occurs around the optimal doping
concentration $x\approx 0.18$, and then decreases in both
underdoped and overdoped regimes, in agreement with the
experiments \cite{tallon}. Within this SC mechanism, we have
calculated DSSF of cuprate superconductors in terms of the
collective mode in the dressed holon particle-particle channel,
and reproduce all main features of inelastic neutron scattering
experiments in the SC-state \cite{dai}, including the energy
dependence of the IC scattering peaks at low energies and
commensurate resonance peak at relatively high energies.

\acknowledgments The author would like to thank Dr. Ying Liang,
Dr. Bin Liu, and Dr. Jihong Qin for the helpful discussions. This
work was supported by the National Natural Science Foundation of
China under Grant No. 10125415, and the Grant from Beijing Normal
University.


\begin{references}
\bibitem [*] {add} To whom correspondence should be addressed.
\end{references}
\end{document}